\documentclass{ws-procs9x6}

\begin{document}

\newcommand{\vs}[1]{\rule[- #1 mm]{0mm}{#1 mm}}
\newcommand{\be}{\begin{equation}}
\newcommand{\ee}{\end{equation}}
\newcommand{\bea}{\begin{eqnarray}}
\newcommand{\eea}{\end{eqnarray}}
\newcommand{\nn}{\nonumber}
\newcommand{\al}{\alpha}
\def\re{{\Re\mbox{e}}}
\def\im{{\Im\mbox{m}}}

\renewcommand{\cite}[1]{[\refcite{#1}]}

\title{Matrix model correlation functions and lattice data for the QCD
  Dirac operator with chemical potential}

\author{\vspace*{-5mm}G. Akemann\footnote{\uppercase{S}upported by a
    \uppercase{DFG} 
    \uppercase{H}eisenberg fellowship and in part by the
    \uppercase{EU} network \uppercase{EUROGRID}.}}

\address
{Service de Physique Th\'eorique, CEA/DSM/SPhT Saclay\\
Unit\'e de recherche associ\'ee au CNRS\\
F-91191 Gif-sur-Yvette C\'edex, France}

\author{T. Wettig\footnote{\uppercase{S}upported in part by
    \uppercase{DOE} grant no.\ \uppercase{DE-FG02-91ER40608}.}} 

\address
{Department of Physics, Yale University, New Haven, CT, 06520-8120, USA\\
RIKEN-BNL Research Center, Upton, NY, 11973-5000, USA}


\maketitle

\vspace*{-5mm}

\abstracts{
We apply a complex chiral random matrix model as an effective model to
QCD with a small chemical potential at zero temperature. 
In our model the correlation functions of complex eigenvalues can be 
determined analytically in two different limits, at weak and strong 
non-Hermiticity. 
We compare them to the distribution of the smallest Dirac operator 
eigenvalues from 
quenched QCD lattice data for small values of the chemical potential,
appropriately rescaled with the volume.
This confirms the existence of two different scaling regimes
from lattice data.
}

\vspace*{-10mm}

\section{Introduction}

The presence of a chemical potential $\mu$ in QCD remains a challenging 
problem, due to the fact that the Dirac operator eigenvalues become complex. 
One possibility to study this situation analytically is given by 
random matrix models. The origin of such an effective description 
for real Dirac eigenvalues at $\mu=0$, as first suggested in 
\cite{Jac}, is by now very well established, and we refer to \cite{JT} for a 
review. A chiral random matrix model including a chemical potential
was proposed in \cite{Steph}. 
While its partition function and susceptibilities 
were determined 
analytically in \cite{Halasz}, the microscopic correlation functions 
are not known to date. Very recently the model has been 
simulated on the lattice \cite{Kostas}, confirming the results 
of \cite{Halasz}. 
In \cite{MPW} complex Dirac spectra have been analyzed using quenched lattice 
data in the bulk of the spectrum. Here chiral symmetry is not important, 
and the relevant 
matrix model correlations are given by the known results for the
Ginibre ensemble \cite{Gin}. 
A transition starting from the Gaussian unitary  
ensemble (GUE) at $\mu=0$ via the Ginibre to the Poisson 
ensemble was observed \cite{MPW}. 

Very recently an alternative chiral matrix model with complex eigenvalues
was introduced and solved in \cite{A02}. Here the microscopic, complex 
correlation functions are calculated in two different limits. 
The first limit is called weakly non-Hermitian and was discovered in 
\cite{FKS}. It interpolates 
between the chiral GUE \cite{Jac} and the correlations in the second limit at 
strong non-Hermiticity. The results of Ginibre \cite{Gin} are at strong 
non-Hermiticity, too, but differ from 
\cite{A02} due to chiral symmetry.
We note that the model \cite{A02} is always in the phase with broken 
chiral symmetry, in contrast to \cite{Steph}.

\section{Chiral random matrix theory with complex eigenvalues}

In this section we summarize the results of the complex extension \cite{A02}
of the chiral GUE. 
The matrix model partition function is defined as  
\bea
Z^{(a)}(\tau) &\equiv& \int\limits_{\mathbb{C}^N}\prod_{j=1}^N
dz_j^2 \ |z_j|^{2a+1}
\exp\!\left[
-\frac{N}{1-\tau^2}\left(|z_j|^2 -\frac{\tau}{2}(z_j^2+z_j^{\ast\,2})\right)
\right]
\nn\\
&&\times \prod_{k>l}^N \left|z_k^2-z_l^2\right|^2.
\label{Zev}
\eea
In analogy to the chiral GUE \cite{Jac}, the power $a=N_f+\nu$ combines 
the number of massless quark flavors $N_f$ and the topological charge $\nu$.
The non-Hermiticity parameter $\tau \in [0,1]$ 
relates to the chemical potential $\mu$ via 
\be
4\mu^2\ =\ 1-\tau^2
\label{mutau}
\ee
from comparing \cite{Steph} and \cite{A02} 
at small $\mu$. 
For $\tau\to1$, the complex eigenvalues 
$z_j$ become real, and we recover the partition function of the 
chiral GUE \cite{Jac}.
 In the limit $\tau\to0$, the non-Hermiticity is maximal, and 
eq.\ (\ref{Zev}) becomes a chiral extension of the Ginibre ensemble \cite{Gin}.
The relation between the matrix model 
 \cite{Steph} and eq.\ (\ref{Zev}) 
consists in
a nontrivial mapping of the complex phase of the 
Dirac determinant in the former to the exponent of the latter,
and we conjecture that the two models are in the same universality class.

The model of 
eq.\ (\ref{Zev}) is solved at finite and large $N$ 
using the technique of orthogonal polynomials in the complex plane \cite{A02}. 
We distinguish two different large-$N$ limits:  
in the weakly non-Hermitian limit \cite{FKS} we keep 
\be
\lim_{N\to\infty}\lim_{\tau\to1}\ 
N(1-\tau^2) \equiv \al^2 
\label{weaklim}
\ee
fixed. We thus send $\mu\to0$ such that the volume $V$ times $\mu^2$ 
is fixed. In this limit 
the eigenvalues macroscopically collapse to the real 
line while microscopically they still extend into the complex plane.
At strong non-Hermiticity we take $N\to\infty$ at fixed $\tau\in[0,1)$. 
In this limit the macroscopic eigenvalue density becomes constant on 
an ellipse in the complex plane. 

The result for the microscopic density at weak non-Hermiticity 
reads\footnote{Note the different normalization 
compared to \cite{A02}.} 
\be
\rho^{(a)}_{\rm weak}(\xi) = \frac{\sqrt{\pi\al^2}}{\mbox{erf}(\al)}\ |\xi|\ 
\exp\!\left[\frac{-1}{\al^2}(\Im m\,\xi)^2\right]
\int_0^1 dt\ \mbox{e}^{-\al^2t} 
J_a(\sqrt{t}\xi)J_a(\sqrt{t}\xi^\ast)\:,
\label{weakrho}
\ee
where we have rescaled the complex eigenvalues 
$\xi \equiv Nz\sqrt{2}$ with the same power in $N$ as  
for real eigenvalues \cite{Jac}. 
In contrast, we obtain for the microscopic density at strong 
non-Hermiticity$^{\mbox{\scriptsize a}}$ 
\be
\rho^{(a)}_{\rm strong}(\xi) = \sqrt{\frac{\pi}{1-\tau^2}}
\ |\xi|\ \exp\!\left[\frac{-1}{2(1-\tau^2)}\,|\xi|^2\right]
I_a\left(\frac{|\xi|^2}{2(1-\tau^2)}\right),
\label{strongrho}
\ee
rescaling $\xi\equiv\sqrt{2N}\,z$ here. 
For the higher order $k$-point eigenvalue correlation functions in both limits
we refer to \cite{A02}. 

\vspace*{-1mm}

\section{Comparison with lattice data}

In this section we compare quenched QCD lattice data with the
predictions of eqs.\
(\ref{weakrho}) and (\ref{strongrho}). The data were generated on a 
$6^4$ lattice at gauge coupling $\beta=5.0$ using staggered fermions. We have 
chosen two different values for the chemical potential, $\mu=0.006$ and 
$\mu=0.2$, with 3500 and 4500 configurations, respectively, 
corresponding to the two different regimes of
weak and strong non-Hermiticity. 
In both cases, we are concentrating on the microscopic scale (i.e.\ a few
eigenvalues) so that instead of unfolding the data we only need to rescale
them by the mean level spacing.
\begin{figure}[hb]
\vspace*{-3mm}
%\epsfxsize=10cm   %width of figure - will enlarge/reduce the figures
%\epsfbox{fig3.eps}
%\figurebox{2cm}{3cm}{} %to have a box alone 
\centerline{\epsfxsize=2.22in\epsfbox{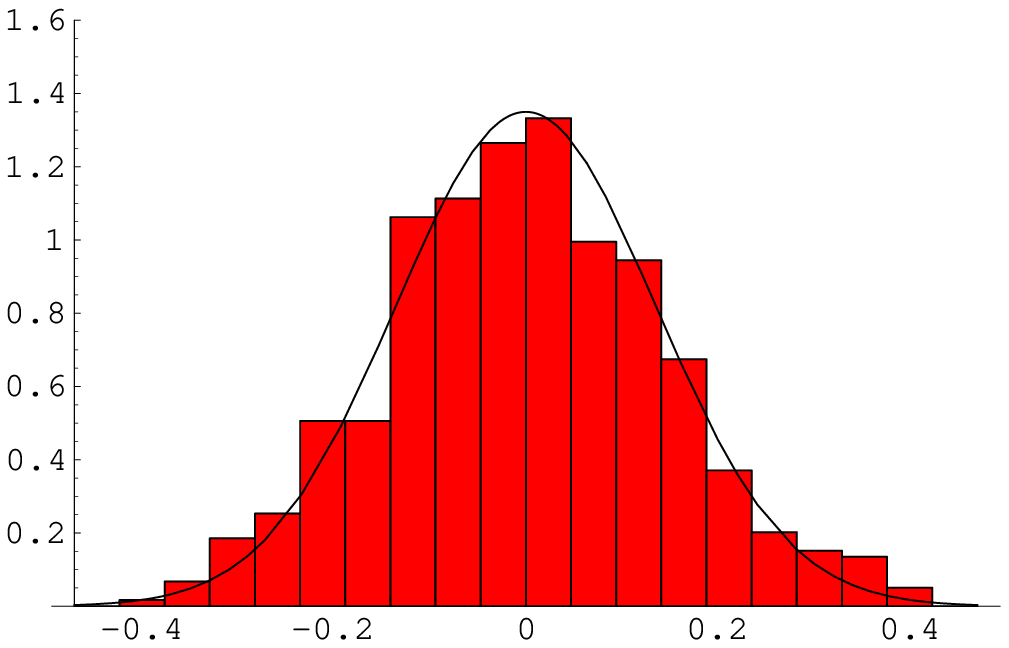}
\epsfxsize=2.22in\epsfbox{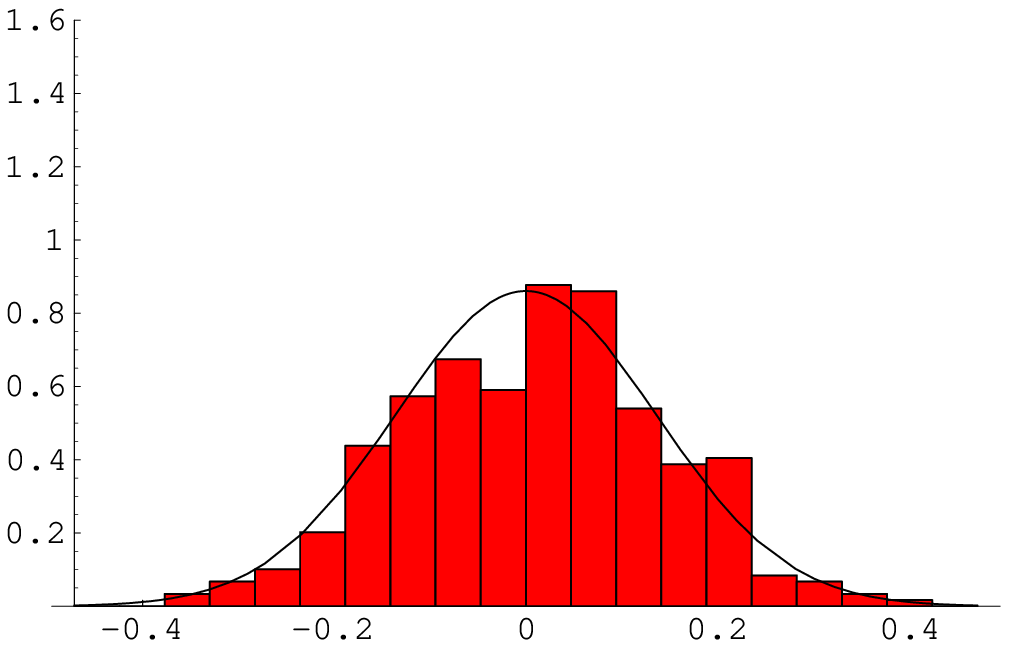}}   
\caption{
  Cuts of the Dirac eigenvalue density for fixed real part of the
  eigenvalues, plot\-ted vs the imaginary part, for $\mu=0.006$, along
  with eq.~(\ref{weakrho}).  Left: real part corresponds to first
  maximum in Fig.~\ref{Recut}, right: real part corresponds to first
  minimum in Fig.~\ref{Recut}.
%  Cuts of the Dirac eigenvalue density for fixed real part of the
%  eigenvalues, plotted versus the imaginary part, for $\mu=0.006$.
%  Left: real part corresponds to the first maximum in Fig.~\ref{Recut},
%  right: real part corresponds to the first minimum in Fig.~\ref{Recut}.
\label{Imcut}}
\end{figure}

For $\mu=0.006$ we determine the parameter 
$\al$ from the weak limit 
by a fit to the decay into the complex plane for a fixed real value, as shown 
in Fig.~\ref{Imcut}.
\begin{figure}[t]
%\epsfxsize=10cm   %width of figure - will enlarge/reduce the figures
%\epsfbox{fig3.eps}
%\figurebox{2cm}{3cm}{} %to have a box alone 
\centerline{\epsfxsize=2.22in\epsfbox{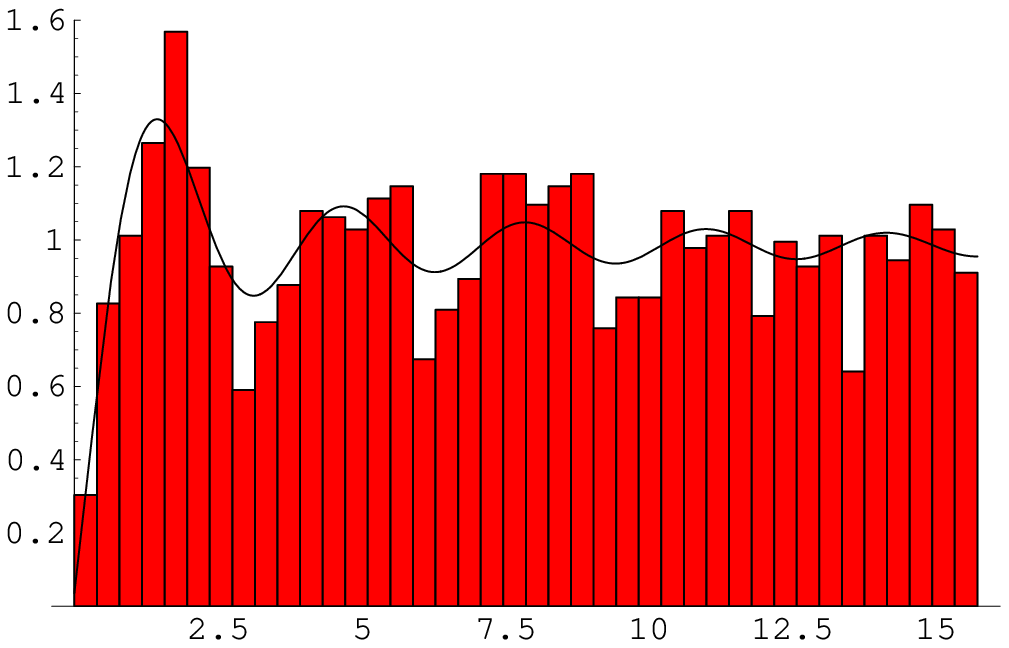}
\epsfxsize=2.22in\epsfbox{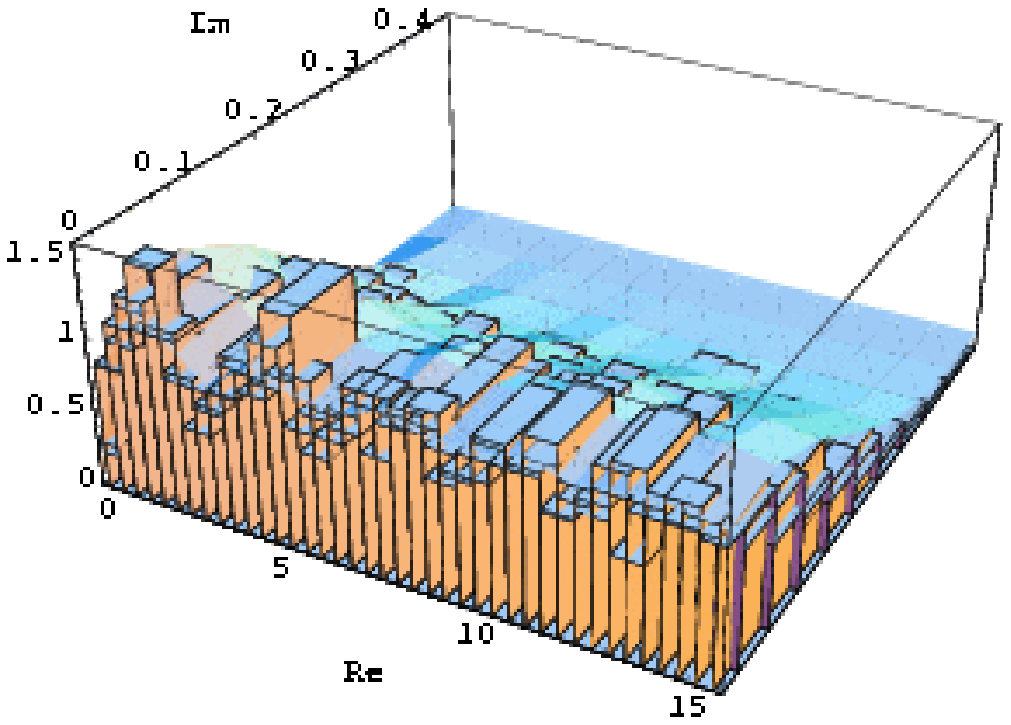}
}   
\caption{
  Left: cut of the Dirac eigenvalue density along the real axis for
  $\mu=0.006$, along with eq.~(\ref{weakrho}).  Right: the density in
  the complex plane, along with eq.~(\ref{weakrho}) (which is slightly
  displaced for comparison).
\label{Recut}}
\vspace*{-2mm}
\end{figure}
We find on average $\al=0.19$, varying about 4\% 
when determined along the first maximum or minimum on the real axis. 
This agrees roughly with 
$\al\sim 0.27$ obtained from  eqs. (\ref{mutau}, \ref{weaklim}) and 
the rescaled level spacing. The discrepancy may be due to the 
uncertainty in the latter and will be further investigated.
In Fig.~\ref{Recut} (left) we show a section of the data along the
real axis. 
Since $\al$ is small, we could equally 
well plot the 
real density \cite{Jac}, 
$\rho^{(a)}_{\rm real}(\xi)= \frac12\pi\xi[J_0(\xi)^2+J_1(\xi)^2]$, 
differing by $\leq 0.5\%$.
The histogram in Fig.~\ref{Recut} (right) shows the
quantitative agreement  
in the complex plane.
\begin{figure}[b]
%\epsfxsize=10cm   %width of figure - will enlarge/reduce the figures
%\epsfbox{fig3.eps}
%\figurebox{2cm}{3cm}{} %to have a box alone 
\vspace*{-2mm}
\centerline{\epsfxsize=2.22in\epsfbox{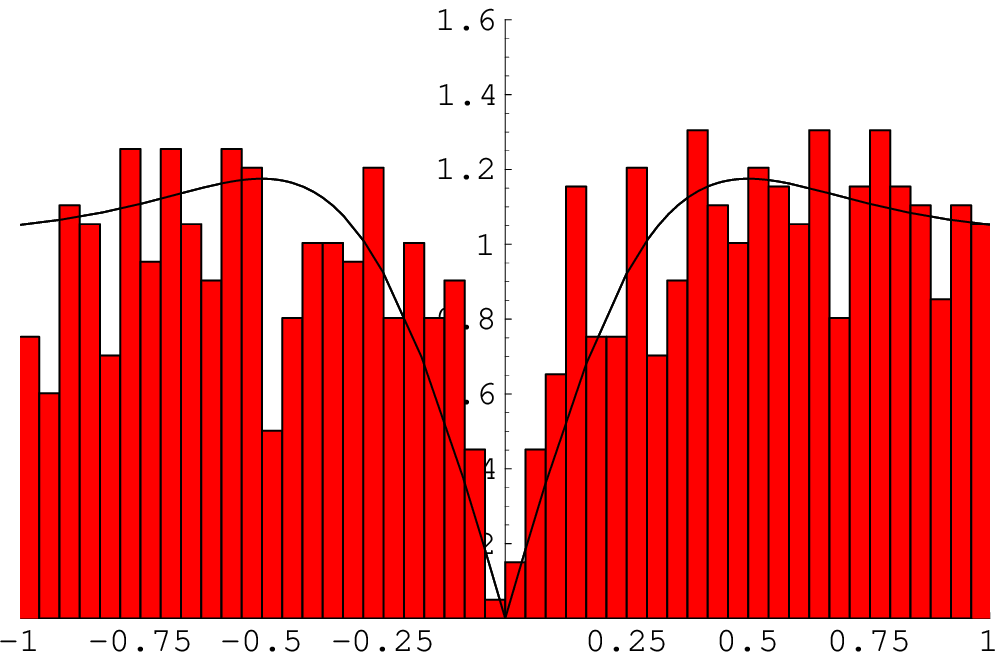}
\epsfxsize=2.22in\epsfbox{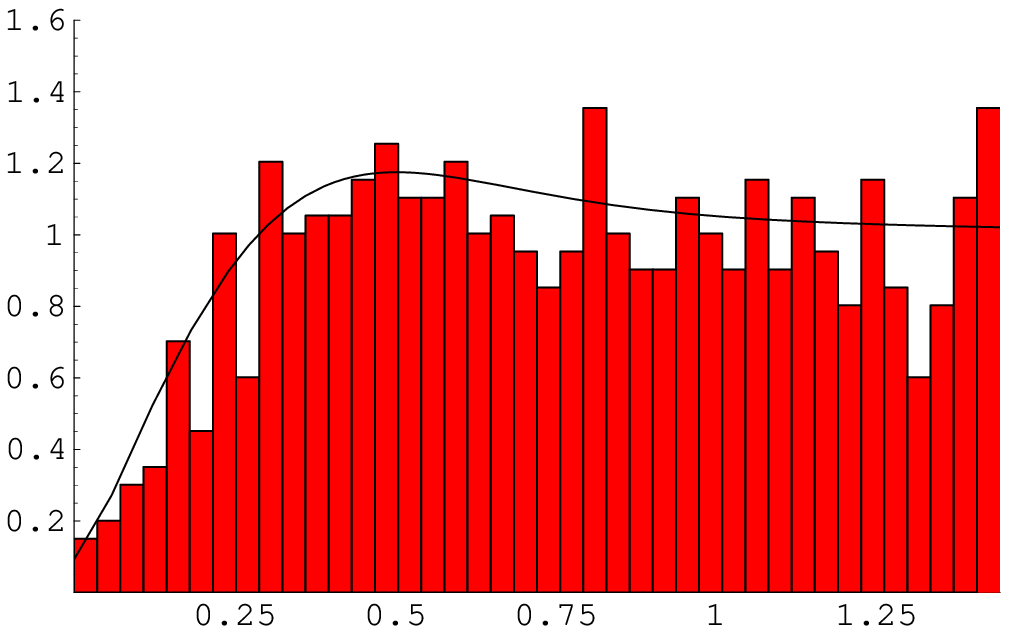}}   
\caption{
  Cuts of the Dirac eigenvalue density for $\mu=0.2$, along with
  eq.~(\ref{strongrho}).  Left: cut along the imaginary axis, right:
  cut along the real axis.
\label{2Dcutstr}}
\end{figure}

\begin{figure}[t]
%\epsfxsize=10cm   %width of figure - will enlarge/reduce the figures
%\epsfbox{fig3.eps}
%\figurebox{2cm}{3cm}{} %to have a box alone 
\centerline{\epsfxsize=2.5in\epsfbox{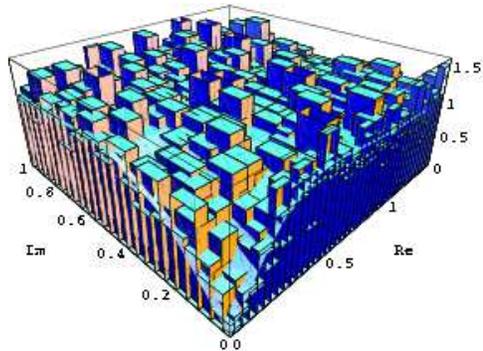}
}   
\caption{
  Dirac eigenvalue density and eq.~(\ref{strongrho}) in the complex
  plane for $\mu=0.2$.
\label{3Dstrong}}
\vspace*{-2mm}
\end{figure}

At strong non-Hermiticity 
we rescale the data for  $\mu=0.2$
independently in the real and imaginary 
directions  with the square root of the respective level spacing.
Cuts along the axes are shown in
Fig.~\ref{2Dcutstr}, and the complete
plot is in Fig~\ref{3Dstrong}.
Here we have used the relation (\ref{mutau}) instead of fitting, 
leading to $\tau=0.9165$. 
The repulsion of the eigenvalues from the origin, which is very different from 
that in the weak limit, is clearly seen in the data.

\vspace*{-1mm}

\section{Conclusions}

In this first exploratory study, we found good agreement between the
predictions of the model (\ref{Zev}) and data from lattice QCD
simulations.  In particular, the two predicted scaling regimes
corresponding to weak and strong non-Hermiticity are clearly visible
in the data.  Obviously, more work is needed to make these conclusions
more quantitative (e.g.\ larger volume, higher statistics, unfolding
details).

\vspace*{-1mm}

\section*{Acknowledgments}
G.A. wishes to thank P. de Forcrand and Z. Fodor for useful discussions.

\end{document}